\documentclass[aps,prb,twocolumn,showpacs,groupedaddress]{revtex4}
\usepackage{amsmath}

\begin{document}

\title{Breakdown of the coexistence of spin-singlet superconductivity and
itinerant ferromagnetism}
\author{R. Shen}
\author{Z. M. Zheng}
\author{S. Liu}
\author{D. Y. Xing}
\affiliation{National Laboratory of Solid State Microstructures, Nanjing
University, Nanjing, 210093 China}

\begin{abstract}
We discuss the possibility of coexistence of spin-singlet superconductivity
and ferromagnetism in a model where the same electrons are assumed
responsible for both of them. Our calculations include both zero and finite
momentum pairing states with both $s$-wave and $d$-wave pairing symmetry.
Under the mean-field approximation, the thermodynamic potential of the
non-magnetic superconducting (SC) state is shown to be always lower than
that of the superconducting ferromagnetic (SF) state. It follows that
the spin-singlet SF state is energetically unfavorable, and a spin-triplet
SF state is more likely to survive in metals such as UGe$_{2}$ and
ZrZn$_{2}$. 
\end{abstract}
\date{\today}


\maketitle

\section{Introduction}
Recently, the observation of superconductivity in ferromagnetic metals,
UGe$_{2}$\cite{uge2}, ZrZn$_{2}$\cite{zrzn2} and URhGe\cite{urhge} has
renewed the interest on the coexistence of ferromagnetism and
superconductivity. The investigation on the superconducting ferrromagnets
traces back to the original works of Clogston\cite{Clogston},
Chandrasekhar\cite{Chandrasekhar}, and Abrikosov \textit{et al.}
\cite{Abrikosov}. At the same time, the possibility of a finite momentum 
pairing state coexisting with the long range ferromagnetic order was
also revealed by Fulde and Ferrell\cite{FF} and by Larkin and
Ovchinnikov\cite{LO}. This finite momentum pairing state is usually called
the FFLO state. These early
works\cite{Clogston,Chandrasekhar,Abrikosov,FF,LO} focused on the
superconductivity in the metals with a spin-exchange field, 
such as produced by ferromagnetically aligned impurities. In such
a superconducting ferromagnet, there are two kinds of electrons,
respectively, responsible for ferromagnetism and superconductivity. One
is the localized electrons forming a ferromagnetic background in the metal
by way of an indirect exchange coupling through itinerant electrons, the
other is the itinerant electrons forming the Cooper pair due to an effective
attractive interaction. Therefore, the magnetic exchange energy $I$ is
independent of the superconducting gap $\Delta $. If the spin-exchange field
is weak enough, the superconductivity can appear against the ferromagnetic
background. The ground state of the system is determined by the ratio
$I/\Delta _{0}$, where $\Delta _{0}$ is the gap in a non-magnetic
superconductor. In the three-dimensional $s$-wave case, for $I/\Delta
_{0}\lesssim 0.707$ all of the itinerant electrons near the Fermi level form
the Cooper pairs with opposite spins and the center-of-mass momentum equal to
zero\cite{Clogston}, while for $0.707\lesssim I/\Delta _{0}\lesssim 0.754$
part of the itinerant electrons near the Fermi level form the Cooper pairs
with a finite center-of-mass momentum and the unpaired electrons show a
finite magnetization which is the paramagnetic response of the itinerant
electrons to the exchange field caused by localized spins\cite{FF,LO}. For
$I/\Delta _{0}\gtrsim 0.754$, the formation of the Cooper pairs is totally
suppressed and the superconductivity is destroyed. In the two-dimensional
$d_{x^{2}-y^{2}}$ superconductor, with the increasing of the exchange field,
the ground state of the system is changed from a zero momentum pairing state
first to a finite momentum pairing state and then to a normal state, the
same as in the $s$-wave case except that the zero momentum pairing
$d_{x^{2}-y^{2}}$ state in the magnetic field has a finite
magnetization\cite{Yang}. 

In the recently
discovered superconducting ferromagnets UGe$_{2}$\cite{uge2} and
ZrZn$_{2}$\cite{zrzn2}, the superconductivity and the ferromagnetism
disappear at the same critical value, $p_{c}$, under the application of the
hydrostatic pressure. This feature may suggest that the same electrons
are responsible for both ferromagnetism and superconductivity in these
novel materials, in contrast to the conventional case of a metal with
magnetic impurities. Very recently, along this 
direction, Karchev \textit{et al.} \cite{Karchev} have developed a
theoretical model, in which the long range ferromagnetic order does not
result from an indirect exchange coupling between the localized
spins, but is a consequence of a spontaneously broken spin rotation symmetry
of those itinerant electrons which participate in the Cooper pair formation.

In this paper, we employ the model in Ref. 10 to discuss
the possibility of the coexistence of spin-singlet superconductivity and
itinerant ferromagnetism. There are two kinds of
interactions between the itinerant electrons in this model. One is an
attractive interaction in the Bardeen-Cooper-Schrieffer (BCS) form resulting
in the superconductivity, the other is an exchange coupling resulting in
the ferromagnetic order. Thus the magnetic exchange energy and the
superconducting gap are related to each other and will be solved
self-consistently. Under the mean-field approximation, the thermodynamic
potential of the superconducting ferromagnetic (SF) state and that of the
non-magnetic superconducting (SC) state are obtained analytically 
at zero temperature. Our calculations include both the $s$-wave and $d$-wave
cases. For each pairing symmetry, both the zero momentum and the finite
momentum pairing states are discussed. It is shown that the thermodynamic
potential of the SF state is always higher than that of the non-magnetic SC
state in all four cases. Therefore, the coexistence of ferromagnetism
and spin-singlet superconductivity can not be realized if the same electrons
are assumed responsible for both of them. A spin-triplet superconductivity is
more likely to survive in those novel superconducting ferromagnets such as
UGe$_{2}$\cite{uge2,Huxley,Machida} and
ZrZn$_{2}$\cite{zrzn2,Fay,Santi,Walker}. In Sec. II to Sec. IV we will
explicitly illustrate this viewpoint for each case and in Sec. V our results
are briefly summarized. 

\section{Zero momentum pairing \lowercase{\textit{s}}-wave case} 
Our starting point is the model Hamiltonian\cite{Karchev}
\begin{eqnarray}
\label {model}
\nonumber
H=\int &d^{3}r& \sum
_{\sigma}c_{\sigma}^{\dagger}(\vec{r})\left(-\frac{1}{2m^{*}}
\vec{\nabla}^{2}-\mu \right)c_{\sigma}(\vec{r})\\
\nonumber
&-&\frac{J}{2}\int d^{3}r\vec{S}(\vec{r})\cdot \vec{S}(\vec{r})\\
&-&g\int d^{3}r c_{\uparrow}^{\dagger}(\vec{r})
c_{\downarrow}^{\dagger}(\vec{r})
c_{\downarrow}(\vec{r})c_{\uparrow}(\vec{r}),
\end{eqnarray}
where $c_{\sigma}(\vec{r})$ are the fermion fields with 
spin $\sigma =\uparrow, \downarrow$ , 
$\vec{S}=\frac{1}{2}\sum _{\sigma, \sigma ' =\uparrow , \downarrow}
c_{\sigma}^{\dagger}\vec{\tau}_{\sigma \sigma '}c_{\sigma '}$ is the spin
field, $\vec{\tau}$ is the Pauli matrices, $m^{*}$ is the effective mass of
the electron, and $\mu$ is the chemical potential. Hamiltonian
(\ref{model}) contains a kinetic energy term, a ferromagnetic
exchange coupling term with strength $J$, and a four-fermion attractive
interaction term with strength $g$. The coupling constant $g$ has a finite
value only in the thin shell of the width $2\epsilon _{c}$ around the Fermi
surface, as in the standard BCS theory. Under the mean-field approximation,
Hamiltonian (\ref{model}) is reduced to\cite{Karchev} 
\begin{eqnarray}
\label {MF}
\nonumber
H_{\text{MF}}&=&\sum _{\vec{p}}\epsilon
_{p}\left(c_{\vec{p}\uparrow}^{\dagger}c_{\vec{p}\uparrow}
+c_{\vec{p}\downarrow}^{\dagger}c_{\vec{p}\downarrow}\right)\\
\nonumber
&-&\frac{JM}{2}\sum_ {\vec{p}}
\left(c_{\vec{p},\downarrow}^{\dagger}c_{\vec{p},\downarrow}-
c_{\vec{p},\uparrow}^{\dagger}c_{\vec{p},\uparrow}\right)+\frac{1}{2}JM^{2}\\
&-&\sum
_{\vec{p}}\left(\Delta
c_{\vec{p},\uparrow}^{\dagger}c_{-\vec{p},\downarrow}^{\dagger}+H.c.\right)
+\frac{|\Delta |^{2}}{g},
\end{eqnarray}
and
\begin{equation}
\label{M}
M=\frac{1}{2}\sum_ {\vec{p}}
\left(\langle c_{\vec{p},\downarrow}^{\dagger}c_{\vec{p},\downarrow}\rangle
-\langle c_{\vec{p},\uparrow}^{\dagger}c_{\vec{p},\uparrow}\rangle \right),
\end{equation}
\begin{equation}
\label{D}
\Delta=\sum _{\vec{p}}g \langle 
c_{-\vec{p},\downarrow}c_{\vec{p},\uparrow}\rangle,
\end{equation}
where $\epsilon _{p}=p^{2}/(2m^{*})-\mu$ is the band energy measured 
from the chemical
potential, $\langle \cdots \rangle $ represents the thermodynamic average,
$M$ defines the magnetization of the system, and $\Delta$ is the
superconducting gap. We note that the two constant terms in Eq. (\ref{MF})
result from the mean-field approximation, $\Delta ^{2}/g$ comes from the BCS
interaction and $JM^{2}/2$ from the exchange coupling. Here the
magnetization defined in Eq. (\ref{M}) arises from a spontaneously breaking
of spin rotation symmetry of the itinerant electrons, which is different
from a paramagnetic response to a magnetic field caused by localized spins.
Therefore, both the gap $\Delta $ and the magnetic exchange energy $I=JM/2$
are determined by Eqs. (\ref{M}) and (\ref{D}) self-consistently, unlike in
the conventional case of a metal with magnetic impurities, where the exchange
energy is considered as an external
parameter\cite{Clogston,Chandrasekhar,Abrikosov,FF,LO,Yang}. 

By means of a Bogoliubov transformation, Hamiltonian (\ref{MF}) can be
diagonalized as 
\begin{equation}
\label{DH}
H_{\text{MF}}=\sum_ {\vec{p}}\left(E_{p}^{\alpha}\alpha _{\vec{p}}^{\dagger}
\alpha _{\vec{p}}+E_{p}^{\beta}\beta _{\vec{p}}^{\dagger}
\beta _{\vec{p}}\right)+E_{0} 
\end{equation}
with 
\begin{equation}
\label{HC}
E_{0}=\sum_ {\vec{p}}\left(\epsilon _{p}-\sqrt{\epsilon
_{p}^{2}+|\Delta |^{2}}\right)+\frac{|\Delta |^{2}}{g}+\frac{JM^{2}}{2},
\end{equation}
where $\alpha _{\vec{p}}$ and $\beta _{\vec{p}}$ are the Bogoliubov fermion
fields with excitation energies
\begin{subequations}
\label{EX}
\begin{eqnarray}
E_{p}^{\alpha}&=&\sqrt{\epsilon _{p}^{2}+|\Delta |^{2}}+I,
\\
E_{p}^{\beta}&=&\sqrt{\epsilon _{p}^{2}+|\Delta |^{2}}-I.
\end{eqnarray}
\end{subequations}
The self-consistent equations (\ref{M}) and (\ref{D}) take the form
\begin{equation}
\label{SE1}
M=\frac{1}{2}\sum _{\vec{p}}\left(n_{p}^{\beta}-n_{p}^{\alpha}\right),
\end{equation}
\begin{equation}
\label{SE2}
|\Delta |=\frac{g|\Delta |}{2}\sum _{\vec{p}}\frac{1-n_{p}^{\alpha}-
n_{p}^{\beta}}{\sqrt{\epsilon _{p}^{2}+|\Delta |^{2}}},
\end{equation}
where $n_{p}^{\alpha , \beta}$ is the momentum distribution function of the
corresponding Bogoliubov fermions. At zero temperature, $n_{p}^{\alpha,
\beta}$ is zero for $E_{p}^{\alpha , \beta}>0$ and one for $E_{p}^{\alpha
, \beta}<0$. The thermodynamic potential at zero temperature can 
be obtained by averaging out the mean-field Hamiltonian (\ref{DH}). We take
the thermodynamic potential of the ideal Fermi gas as the origin of the
energy in the following discussions. 

Obviously, the self-consistent Eqs. (\ref{SE1}) and (\ref{SE2})
have a non-magnetic SC solution with a finite gap $\Delta _{0}$ and zero
magnetization. By replacing the summation over the momentum $\vec{p}$ in Eq.
(\ref{SE2}) with an integral over $\epsilon _{p}$, for the weak coupling
limit $gN(0)\ll 1$,  one can easily find that the energy gap in the
non-magnetic SC solution is given by 
\begin{equation}
\label{D0}
\Delta _{0}=2\epsilon _{c} \exp \left(-\frac{1}{gN(0)}\right),
\end{equation}
where $N(0)$ is the density of states (DOS) at the Fermi level. 
Therefore, the thermodynamic potential of the non-magnetic SC state at zero
temperature takes the well-known form 
\begin{equation}
\Omega _{\text{SC}}=-\frac{1}{2}N(0)\Delta _{0}^{2}.
\end{equation}

Next, we discuss the SF solution at zero temperature with both finite gap
and finite magnetization. Here, the gap is a real number, and the
self-consistent Eqs. (\ref{SE1}) and (\ref{SE2}) turn to
\begin{equation}
\label{SE3}
M=N(0)\int_{0}^{\sqrt{I^{2}-\Delta ^{2}}}d\epsilon,
\end{equation}
\begin{equation}
\label{SE4}
\frac{1}{gN(0)}=\ln{\frac{2\epsilon _{c}}{\Delta _{0}}}
=\int _{\sqrt{I^{2}-\Delta ^{2}}}^{\epsilon _{c}}
\frac{d\epsilon}{\sqrt{\epsilon ^{2}+\Delta ^{2}}}.
\end{equation}
The integrals over $\epsilon _{p}$ in the self-consistent
equations are confined in a very thin shell around the Fermi level, and
therefore we have replaced the DOS in the integrals by $N(0)$. It is shown
that, in the SF state, only part of the itinerant electrons whose energies
are in the range $\sqrt{I^{2}-\Delta ^{2}}<|\epsilon |<\epsilon _{c}$ form
the Cooper pairs while other electrons with energies in the range
$0<|\epsilon |<\sqrt{I^{2}-\Delta ^{2}}$ remain unpaired. These unpaired
itinerant electrons give rise to a finite spontaneous magnetization.
Completing the integrals in Eqs. (\ref{SE3}) and (\ref{SE4}), one finds
that, for $r>1$, the SF solution is given by 
\begin{equation}
\label{SFS}
I=\frac{r}{\sqrt{r^{2}-1}}\Delta ,
\;\;\;
\Delta =\sqrt{\frac{r-1}{r+1}}\Delta _{0},
\end{equation}
where $r=JN(0)/2$ is the dimensionless measurement for the exchange
coupling strength. Part of the electrons near the Fermi level remain
unpaired so that the superconducting gap $\Delta $ in the SF
state is always less than $\Delta _{0}$. Here, the exchange energy $I$
results from the exchange coupling between the itinerant electrons and is
related to the gap $\Delta $, in contrast to the case of a metal with
impurities. For fixed $r$, the ratio $I/\Delta $ is a constant, independent
of $\Delta$. When the gap vanishes the exchange energy also decreases to
zero. Substituting the SF solutions for the magnetization 
and the gap into Eqs. (\ref{DH}) and (\ref{HC}) and
averaging out the mean-field Hamiltonian (\ref{DH}), one obtains the
thermodynamic potential of the SF state at zero temperature as 
\begin{equation}
\label{O1}
\Omega _{\text{SF}}=-\frac{1}{2}N(0)\Delta ^{2}.
\end{equation}
Comparing $\Omega _{\text{SC}}$ with $\Omega _{\text{SF}}$, one
finds that the SF state in the zero momentum pairing $s$-wave case is always
energetically unfavorable due to $\Delta < \Delta _{0}$. This result can be
understood by the following argument. The spontaneous magnetization in the
SF state is very weak and the exchange energy $I$ in the SF state is always
less than $\Delta _{0}$. From Eq. (\ref{MF}) and (\ref{SFS}) one finds that
the Zeeman energy gained by the unpaired electrons can not compensate for
the loss of the condensate energy in depairing the Cooper pairs. Therefore,
all the itinerant electrons near the Fermi level remain paired with opposite
spins, and the non-magnetic SC state is always more stable than the SF
state. This feature is similar to that in the conventional case of a metal
with impurities, where for $I/\Delta _{0}\lesssim 0.707$ the itinerant
electrons form a non-magnetic SC state in the presence of an exchange field
of the ferromagnetic background. In the present case, the external exchange
field is absent, so is the spontaneous magnetization of the itinerant
electrons. As a result, the coexistence of $s$-wave superconductivity with
zero momentum pairing and itinerant ferromagnetism can not be realized.

\section{Finite momentum pairing \lowercase{\textit{s}}-wave case}
In the conventional case of a metal with impurities, the formation of a
condensate of finite center-of-mass momentum turns out to be more
advantageous\cite{FF,LO}. Such a finite momentum pairing state, where part
of the itinerant electrons near the Fermi level form the Cooper pairs with a
finite center-of-mass momentum and the unpaired electrons show a finite
magnetization, is the ground state of the system when the exchange energy
$I$ is in a proper range. However, if the exchange energy is caused by the
spontaneous magnetization of the itinerant electrons and thus related
to the gap, the conclusion is completely different, as will be shown below.

Here, the center-of-mass momentum of the Cooper pair, $\vec{q}$, is not
equal to zero, and thus the superconducting gap turns out to be a periodic
function of the coordinates\cite{LO}, such as $\Delta
(\vec{r})=\Delta \exp{(i\vec{q}\cdot \vec{r})}$. The excitation energies of
the Bogoliubov fermions in such finite momentum pairing state are given
by\cite{Abrikosov1} 
\begin{subequations}
\label{EXP}
\begin{eqnarray}
E_{p}^{\alpha}&=&\sqrt{\epsilon _{p,q}^{2}+\Delta ^{2}}+I
+\frac{1}{2}v_{F}q\cos {\theta},
\\
E_{p}^{\beta}&=&\sqrt{\epsilon _{p,q}^{2}+\Delta ^{2}}-I
-\frac{1}{2}v_{F}q\cos {\theta},
\end{eqnarray}
\end{subequations}
where $\epsilon _{p,q}=(\epsilon _{\vec{p}+\vec{q}/2}+\epsilon
_{-\vec{p}+\vec{q}/2})/2=(p^{2}+q^{2})/(2m^{*})$, $v_{F}$ is the Fermi
velocity, and $\theta$ is the angle between the momentum $\vec{p}$
and $\vec{q}$. The diagonalized Hamiltonian, the self-consistent equation of
the magnetization and the gap equation have the same forms as Eqs.
(\ref{DH}), (\ref{HC}), (\ref{SE1}), and (\ref{SE2}), respectively, provided
that $\epsilon _{p}$ is replaced by $\epsilon _{p,q}$ and the excitation
energies in Eqs. (\ref{EX}a) and (\ref{EX}b) are replaced by those in Eqs.
(\ref{EXP}a) and (\ref{EXP}b). The momentum $q$ is determined by the
minimization of the thermodynamic potential, and $v_{F}q/2$ is at most of
the order $\Delta _{0}$\cite{FF}. By replacing the summation over the
momentum $\vec{p}$ with an integral over $\epsilon =\epsilon _{p,q}/\Delta
$, the self-consistent equations at zero temperature can be written as
\begin{eqnarray}
\label{I1}
\nonumber
\ln{\frac{\Delta}{\Delta _{0}}}&=&
-\frac{1}{4}\left(\int_{E_{p}^{\alpha}<0}
\frac{dxd\epsilon}{\sqrt{\epsilon
^{2}+1}}+\int_{E_{p}^{\beta}<0}\frac{dxd\epsilon}{\sqrt{\epsilon
^{2}+1}}\right)
\\
&=&-F_{1}(I',y),
\end{eqnarray}
\begin{equation}
\label{I2}
I'=\frac{1}{4}r\left(\int_{E_{p}^{\beta}<0} dxd\epsilon
-\int_{E_{p}^{\alpha}<0}dxd\epsilon \right)
=rF_{2}(I',y),
\end{equation}
\begin {equation}
\label{I3}
\frac{1}{N(0)\Delta ^{2}}
\frac{\partial \Omega}{\partial y}=F_{3}(I',y)=0,
\end{equation}
where $I'=I/\Delta $, $y=(v_{F}q/2)/\Delta $, $x=\cos {\theta}$ and
$F_{1,2,3}$ are functions of $I'$ and $y$, independent of $\Delta$. From
Eqs. (\ref{I2}) and (\ref{I3}), one finds that $I'$ and $y$ 
are determined only by $r$. Hence, the ratio $\Delta /\Delta _{0}$ in Eq.
(\ref{I1}) is also a constant for fixed $r$. Combining with Eq. (\ref{D0}),
one obtains 
\begin{equation}
\frac{d(\frac{1}{g})}{d\Delta}=-N(0)\frac{1}{\Delta}.
\end{equation}
This result is quite different from
that in the conventional case where $I$ is an external parameter and the
right-hand side of Eq. (\ref{I1}), for fixed $I$, depends on the gap
explicitly. The thermodynamic potential of the SF state with moving pairs
is given by applying the formula\cite{Abrikosov1}
\begin{equation}
\label{Feynman}
\Omega _{\text{SF}}-\Omega (\Delta =0)=\int _{0}^{\Delta}\Delta
^{2}\frac{d(\frac{1}{g})}{d\Delta}d\Delta , 
\end{equation}
where $\Omega (\Delta =0)$ is the energy when the gap vanishes. In our
cases, the ratio $I'=I/\Delta $ is a constant for fixed $r$. When the gap
vanishes the exchange energy also decreases to zero. Therefore, $\Omega
(\Delta =0)$ here is actually the energy of the ideal Fermi
gas and is set to zero in our discussions. Completing the integral in 
Eq. (\ref{Feynman}), one obtains 
\begin{equation}
\label{O2}
\Omega _{\text{SF}}=-\frac{1}{2}N(0)\Delta ^{2}.
\end{equation}
The energy in the SF state with moving pairs defined in Eq. ({\ref {O2}})
takes the same form as that with immobile pairs defined in Eq. ({\ref {O1}),
but the gap functions in the two cases are different, because they are
determined by two different self-consistent equations. Noting that the gap
$\Delta $ is always less than $\Delta _{0}$, we find that the thermodynamic
potential of the SF state with moving pairs is still higher than that of the
non-magnetic SC state. Our result is different from that in FFLO state, in
which part of the itinerant electrons near the Fermi level remain unpaired
and give rise to a finite magnetization in an exchange field of
$0.707\lesssim I/\Delta _{0}\lesssim 0.754$ and other
electrons form the Cooper pairs with a finite center-of-mass
momentum\cite{FF}. This is because that the Zeeman energy gained by the
unpaired electrons in the SF state in the present model is less than that in
FFLO state. In the conventional case, the exchange
field is caused by localized spins and the exchange energy $I$ is an
external parameter. Each itinerant electron moving in the exchange field has
an additional energy $I$ or $-I$ depending on the spin orientation of the
electron. The difference in number between spin-up and spin-down electrons
results in a finite magnetization $M$. The Zeeman energy of the system is
given by $-2IM$. In the present model, the exchange field is due to the
spontaneous magnetization of the itinerant electrons and the exchange energy
$I$ is determined self-consistently. Therefore, the additional energy $I$
gained by the electron in the exchange field here is actually the
interaction energy between this electron and other itinerant electrons. The
Zeeman energy of the system is obtained by adding the exchange energy of
every electron together, but the interaction energy between the two
electrons should be countered only once. Thus, the Zeeman energy here is
$-IM$, which is only a half of that in an external field with the same $I$.
This feature is embodied in Hamiltonian (\ref{MF}) via the constant term
$JM^{2}/2$. Such a loss in the Zeeman energy results that the SF state with
moving pairs is also energetically unfavorable and the coexistence of the
itinerant ferromagnetism and the superconductivity is more difficult. As a
result, if the exchange energy is caused by the spontaneous magnetization of
the itinerant electrons and determined self-consistently, the $s$-wave
non-magnetic SC state is always more stable than the $s$-wave SF state, no
matter whether the center-of-mass momentum of the Cooper pair is zero or
not.

\section{\lowercase{\textit{d}}-wave case}
\subsection{Zero momentum pairing \textit{d}-wave case}
We take a two-dimensional $d_{x^{2}-y{2}}$ superconductor for example. The
gap function takes the form $\Delta (\theta )=\Delta _{\text{d}} \cos
(2\theta) $, where $\theta $ is the azimuthal angle of momentum
$\vec{p}$. The self-consistent equation for the magnetization remains the
form of Eq. (\ref{SE1}) and the gap equation is given by\cite{Yang} 
\begin{equation}
\label{SED}
\Delta _{\text{d}}=\frac{g\Delta _{\text{d}}}{2}\sum
_{\vec{p}}\frac{1-n_{p}^{\alpha}- n_{p}^{\beta}}{\sqrt{\epsilon
_{p}^{2}+\Delta _{\text{d}}^{2}\cos ^{2}(2\theta )}} \cos ^{2}(2\theta ),
\end{equation}
where $n_{p}^{\alpha}$ and $n_{p}^{\beta}$ are the momentum distribution
functions with the energies
\begin{subequations}
\begin{eqnarray}
E_{p}^{\alpha}&=&\sqrt{\epsilon _{p}^{2}+\Delta _{\text{d}}^{2}\cos
^{2}(2\theta )}+I, 
\\
E_{p}^{\beta}&=&\sqrt{\epsilon _{p}^{2}+\Delta _{\text{d}}^{2}\cos
^{2}(2\theta )}-I. 
\end{eqnarray}
\end{subequations}

At first, we consider the non-magnetic SC solution where $I=0$ and thus
$n_{p}^{\alpha}=n_{p}^{\beta}=0$. Carrying out the summation in Eq.
(\ref{SED}), the gap in the non-magnetic SC state is obtained by 
\begin{equation}
\label{ND}
\Delta _{\text{d}0}=2.43\epsilon _{c} \exp (-\frac{2}{gN(0)}).
\end{equation}
By using Eq. (\ref{Feynman}), the thermodynamic potential of
the $d_{x^{2}-y{2}}$ non-magnetic SC state at zero temperature is given by
\begin{equation}
\Omega _{\text{SC}}=-\frac{1}{4}N(0)\Delta _{\text{d}0}^{2}.
\end{equation}

Next, we turn to the SF solution, where only part of the itinerant
electrons near the Fermi surface form the Cooper pairs while other electrons
whose energies satisfy the inequality, $E_{p}^{\beta}<0$, remain unpaired.
Replacing the summations over momentum $\vec{p}$ in Eqs. (\ref{SE1}) and
(\ref{SED}) by the integrals over the energy $\epsilon=\epsilon _{p}/\Delta
_{\text{d}}$ and the azimuthal angle $\theta$, the gap equation and the
self-consistent equation of the magnetization are reduced to 
\begin{equation}
\label{DS1}
\ln{\frac{\Delta _{\text{d}}}{\Delta _{\text{d}0}}}=
-\frac{1}{2\pi}\int _{E_{p}^{\beta}<0}d\theta d\epsilon 
\frac{\cos ^{2}(2\theta )}
{\sqrt{\epsilon ^{2}+\cos ^{2}(2\theta )}}
\end{equation}
\begin{equation}
\label{DS2}
\frac{I}{\Delta _{\text{d}}}=\frac{1}{4\pi}r\int _{E_{p}^{\beta}<0}d\theta
d\epsilon . 
\end{equation}
From Eq. (\ref{DS2}), it follows that the ratio $I/ \Delta _{\text{d}}$ is a
constant for fixed $r$, and therefore the ratio $\Delta _{\text{d}}/\Delta
_{\text{d}0}$ in Eq. (\ref{DS1}) is also a constant for fixed $r$.
The exchange energy $I$ is related to the gap $\Delta _{\text{d}}$, and
decreases to zero as the gap vanishes. With the help of Eqs.
({\ref{Feynman}) and (\ref{ND}), the thermodynamic potential of SF state is
given by 
\begin{equation}
\label{O3}
\Omega _{\text{SF}}=-\frac{1}{4}N(0)\Delta _{\text{d}}^{2}.
\end{equation}
The gap in the SF state is always less than that in the non-magnetic SC
state. Hence, the $d_{x^{2}-y{2}}$ SF state always has higher energy than
the $d_{x^{2}-y{2}}$ non-magnetic SC state. This result is the same as that
in the s-wave case. 

\subsection{Finite momentum pairing $d$-wave case}
In the $d_{x^{2}-y{2}}$ SF state with finite momentum pairing, the
excitation energies of the quasi-particles take the form
$E_{p}^{\alpha}=\sqrt{\epsilon _{p,q}^{2}+\Delta _{\text{d}}^{2}\cos
^{2}(2\theta )}+I+(v_{F}q/2)\cos (\theta -\theta _{q})$, and
$E_{p}^{\beta}=\sqrt{\epsilon _{p,q}^{2}+\Delta _{\text{d}}^{2}\cos
^{2}(2\theta )}-I-(v_{F}q/2)\cos (\theta -\theta _{q})$, where $\theta _{q}$
is the azimuthal angle of the momentum $\vec{q}$ which is determined by the
minimization of the thermodynamic potential. The self-consistent equations
in the finite momentum pairing $d$-wave case have the same structure as that
in the $s$-wave case. The equations for the magnetization and the momentum
of the pair can be reduced to three expressions which only depend on
$I/\Delta _{\text{d}}$, $(v_{F}q/2)/\Delta _{\text{d}}$, $\theta _{q}$ and
$r$. Therefore, the variables $I/\Delta _{\text{d}}$, $(v_{F}q/2)/\Delta
_{\text{d}}$ and $\theta _{q}$ are found to be constant for fixed $r$. The
gap equation is reduced to the expression which only contains the variables
$\Delta _{\text{d}}/\Delta _{\text{d}0}$, $I/\Delta _{\text{d}}$,
$(v_{F}q/2)/\Delta _{\text{d}}$, and $\theta _{q}$. Thus, the ratio $\Delta
_{\text{d}}/\Delta _{\text{d}0}$ is also a constant for fixed $r$. Combined
with Eqs. (\ref{ND}) and (\ref{Feynman}), the thermodynamic potential of the
SF state with finite momentum pairing is obtained as the same form as
defined by Eq. (\ref{O3}) but the gap here is different from that in the
zero momentum pairing case. It then follows that the energy in the
$d_{x^{2}-y{2}}$ SF state with moving pairs is also higher than that of the
non-magnetic SC state due to $\Delta _{\text{d}}<\Delta _{\text{d}0}$ and
therefore the SF state can not be realized. 

\section{Summary}
In this paper, we employ a model including both an attractive interaction
and an exchange coupling between the itinerant electrons to study the
possibility of the coexistence of spin-singlet superconductivity and
itinerant ferromagnetism. In this model the electrons responsible for the
ferromagnetism and those forming the Cooper pairs are the same and thus the
exchange energy between the two spin subbands is determined
self-consistently, as opposed to the conventional case where the exchange
energy is an external parameter. Under the mean-field approximation, the
self-consistent equations of both the superconducting gap and the exchange
energy are considered simultaneously, and the thermodynamic potential in the
non-magnetic SC solution and that in the SF solution are obtained
analytically at zero temperature. We discussed four cases including both the
zero and finite momentum pairing state with both the $s$-wave and $d$-wave
pairing symmetry. One finds that the exchange energy and the superconducting
gap in the SF solution are related to each other. It is shown that the
Zeeman energy gained by the unpaired electrons in the SF state can not
compensate for the loss of the condensate energy in depairing the Cooper
pairs and thus the thermodynamic potential of the SF state is always higher
than that of the non-magnetic SC state in both the $s$-wave case and the
$d$-wave case, no matter whether center-of-mass momentum of the pair is zero
or not. Therefore, the coexistence of ferromagnetism and spin-singlet
superconductivity can not be realized if the same electrons are assumed
responsible for both of them. The present results indicate that a
spin-triplet superconductivity is more likely to survive in those novel
superconducting ferromagnets such as UGe$_{2}$ and ZrZn$_{2}$, which are
consistent with some new progress\cite{Huxley,Machida,Santi,Walker}. 



\end{document}